# Accurate decoding of materials using a finger mounted accelerometer

Kuniharu Sakurada, Gowrishankar Ganesh, Wenwei Yu, and Kahori Kita

*Abstract*— Sensory feedback is the fundamental driving force behind motor control and learning. However, the technology for low-cost and efficient sensory feedback remains a big challenge during stroke rehabilitation, and for prosthetic designs. Here we show that a low-cost accelerometer mounted on the finger can provide accurate decoding of many daily life materials during touch. We first designed a customized touch analysis system that allowed us to present different materials for touch by human participants, while controlling for the contact force and touch speed. Then, we collected data from six participants, who touched seven daily life materials-plastic, cork, wool, aluminum, paper, denim, cotton. We use linear sparse logistic regression and show that the materials can be classified from accelerometer recordings with an accuracy of 88% across materials and participants within 7 seconds of touch.

## I. INTRODUCTION

Sensory feedback is crucial for human motor control and learning [1]. Sensory feedback also modulates the usage of limbs, both in the healthy [2], and especially in stroke patients [3]. Finally, tactile sensory feedback is well known to be crucial for gripping tasks [4, 5].For these above reasons, researchers have strived to include sensory feedback modalities in rehabilitation of stroke [5, 6] and in the design of prosthesis [4]. One would assume that stimulating sensory nerves directly can provide rich and 'natural' sensory information, but this procedure is invasive and yet to be perfected [7]. Similarly, external (non-invasive) sensory feedback systems, for example [5, 8], that proposed to provide a transcutaneous electrical stimulation modulated by the amount of grip force are to be developed for feedback of texture and material properties [9]. Overall, a sensory feedback system that can provide information about texture as well as force to users with sensory deficits, and amputees is still absent. Our aim is to develop a low-cost and portable external (non-invasive) sensory feedback system which can address this issue by detecting material texture and contact force when an object is touched, and then providing a corresponding feedback to the user on an unaffected limb. In this study, we concentrate on the first part of our aim- that of develop a low-cost sensory

*Research was supported by Grants-in-Aid for Young Scientists (A) (15H05357) to KK.

K. Sakurada is with the Medical Engineering Department, Chiba University, Inage, Chiba 2638522 Japan (email: afpa5523@chiba-u.jp).
G. Ganesh is with CNRS-AIST JRL UMI3218/RL 1-1-1 Umezono, 305-8560 Tsukuba, Japan, and CNRS-University of Montpellier, LIRMM Interactive Digital Human group, 161 Rue Ada, 34095 Montpellier Cedex 5, France. (e-mail: gans_gs@hotmail.com)
W. Yu is with Center for Frontier Medical Engineering, Chiba University, Inage, Chiba 2638522 Japan (e-mail: yuwill@faculty.chiba-u.jp).
K. Kita is with Center for Frontier Medical Engineering, Chiba University, Inage, Chiba 2638522 Japan (corresponding author, phone: +81- 43-290-3499; e-mail: kkita@chiba-u.jp).

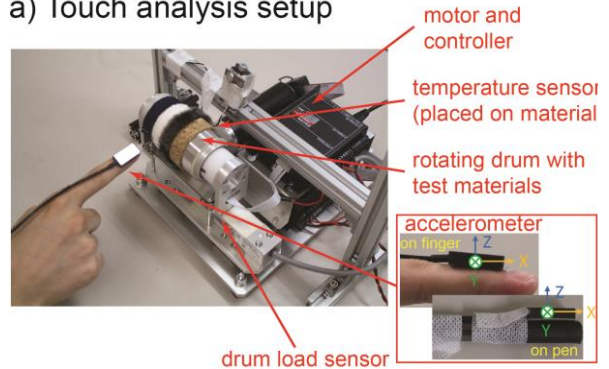
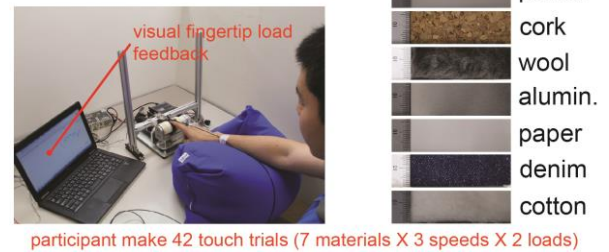

Figure 1. a) Our custom made touch analysis setup includes a rotating drum on which we can mount upto 5 materials at one time. The drum includes a load force sensor and a temperature sensor. (b) Our experiment required participants to touch seven different materials with (see right panel of Fig. 1B) presented at different speeds. The participants maintained a predefined load force by utilizing the force feedback presented on the screen in front of them.

detection (decoding) system that can recognize materials encountered in daily life.

Contact with any surface is associated with movement which is expected to lead to small vibrations of the finger. Previous studies have shown that these vibrations can be useful to detect roughness of touched materials [15,16].Using a stethoscope, Delhay et al. have showed that vibrations generated as a result of interaction of a finger with a rough surface is transmitted even uptill the wrist [10]. We therefore hypothesized that an accelerometer mounted on a finger may serve as a suitable transducer to decode surface texture, and hence material. State of art accelerometers are small, light and cheap to buy. A previous study by Fagiani et al. 2011 has shown that accelerometers can detect differences in movement speed and roughness [xxx]. However, it was unclear whether the vibrations induced by a touched object, as detected by a finger mounted accelerometer, are rich enough to decode various textures. To anticipate the result of our study, we show that accelerometer detected vibrations can be rich enough to differentiate many daily life materials independent of the speed and load characteristics of the touch- we decode seven materials in this study.

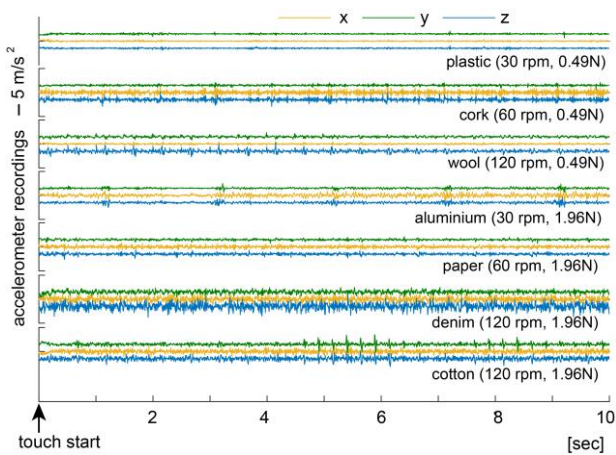

Figure 2. The figure shows the sample accelerometer readings from one representative participant recorded from when he touched the seven materials in different trials.

To investigate whether and how a touched surface can be decoded using accelerometer signals, we first developed a customized touch analysis setup that allows us to present different materials to be touched by human participants, at predefined contact forces, and speeds. With this device we recorded accelerometer data when participants touched seven daily life materials (specifically cotton, plastic, cork, wool, aluminum, paper, denim) with their finger, with two different contact forces, and three contact speeds. We also analyzed the case when the participants touched the materials with a rubber tipped plastic pen (with the accelerometer on the pen in this case) as a first simulation of touch with a prosthetic finger. We then analyzed the accelerometer signals using sparse logistic regression (SLR) [9]. To anticipate our results, we show that the seven materials can be identified with an accuracy of over 80% within 5 seconds of the start of the touch using an accelerometer affixed to a real finger, and with 100% accuracy for five of the seven materials with an accelerometer on a simulated prosthetic finger.

## II. METHODS

### A. Experimental setup

Our custom made touch analysis system is shown in Fig. 1a. The texture is presented to the participants on a rotating drum which we can set up to five surfaces at one time (Fig. 1a). The rotation speed of the drum is controlled by a DC motor (A-max 22, 110164, Maxon, gearbox from Planetary Gearhead, GP22C, 143980, Maxon). A single axis single point type load cell (PW6C, UNIPULSE) on the drum base allows us to measure the normal load force on the drum. A temperature sensor (MLT422/A Skin Temperature Probe, Thermistor Pod, AD Instruments) allows us to measure surface temperature of the surface of test materials. Data from this sensor is not presented in the current study.

Six healthy participants (all men, mean age: 22.6 ± 0.5 years, all right-handed) were asked to touch each of seven test materials (plastic, cork, wool, aluminum, paper, denim and cotton, 15 mm width) rotating at a constant speed (30, 60, or 120 rpm) for 10 s while maintaining a constant load force (0.49, 1.96 N). See Fig. 1b. The participants maintained the load force aided by a visual feedback from the force sensor (passed through a digital converter from Power Lab 16/35, AD Instruments), and displayed on the force feedback monitor in front of them (Lab Chart, AD Instruments) in real time. The participants touched the presented material/s with their index finger. Each participant touched every test material at all the (three) speeds and (two) contact loads. Therefore each participant worked in 3 (speeds)X 2 (loads) X 7 (materials) =42 trials in total.

An accelerometer (six axes = x, y, z, roll, pitch, yaw, MP-M6-06/2000C, Micro-Stone Corporation) was attached to the right index finger (on the second phalange) to measure vibration information of the fingertip (Fig.1a, inset). The data sampling from the accelerometer and load cell were all performed at 200 Hz.

The same procedure as above was used for the simulated prosthetic finger (the plastic pen). The acceleration sensor was attached to the top of the pen, above the rubber cover (see Fig. 1a, inset) while it as held by an experimenter, who touched on the test materials with it while maintaining the required load force.

### B. Data preprocessing for texture decoders

The acceleration signal from each axis was high-pass filtered above 1 Hz. We used both frequency and time features of the acceleration. We considered 150 ms time bins and extracted the median power every 10 Hz between 0 to 100 Hz as frequency features. We calculated the root mean square (RMS) acceleration every 150 ms as the first amplitude feature. Finally, we took the average first order difference signal every 150 ms as the second amplitude feature. As our accelerometer gives a 6-dimensional signal, this gave us 10* 6 =60 frequency features and 2*6 =12 amplitude features, that is 72 features in total every 150 ms.

In our study, we chose to use a sparse logistic regression (SLR) decoder [11] due to the good performance observed with this decoder in other works by the authors [12, 13]. We constructed a one-vs-rest decoder for each material, so as to have seven material decoders in total. Each material decoder classified a data bin as belonging to that material or not. For example, the 'plastic vs other' decoder performs a binary decoding to decide whether a data bin in a trial comes from during touching of plastic or not, the 'cotton vs other' performs a binary decoding to decide whether a data bin in a trial comes from during touching of cotton or not, and so on.

As mentioned earlier, each trial consisted of a touch of 10 seconds by the participants. We were interested in looking at the decoding performance with time, i.e to see how the decoding accuracy changes with the length of touch, and therefore adopted the following procedure to test the decoding performance.

We trained the decoder for every individual with the last 3 seconds of accelerometer data from the 42 trials. As we analyzed the data in 150 ms bins, this procedure resulted in 42X20=840 training data points from each participant, which was used to train the seven material decoders for the participant. The trained decoders were tested over 42 validations, using the first seven seconds of (untrained) data from each trial. The testing was done six times, using data between 1 and 2 seconds after the start of touch (i.e six 150 ms bins), between 1 and 3 seconds after start of touch (13 150ms bins), between 1 and 4 seconds after start of touch (20 bins), between 1 and 5 seconds after start of touch (26 bins), between 1 and 6 seconds after start of touch (33 bins) and between 1 and 7 seconds after start of touch (40 bins). We left out the first 1 second of data to avoid the transient behavior at the start of touch.

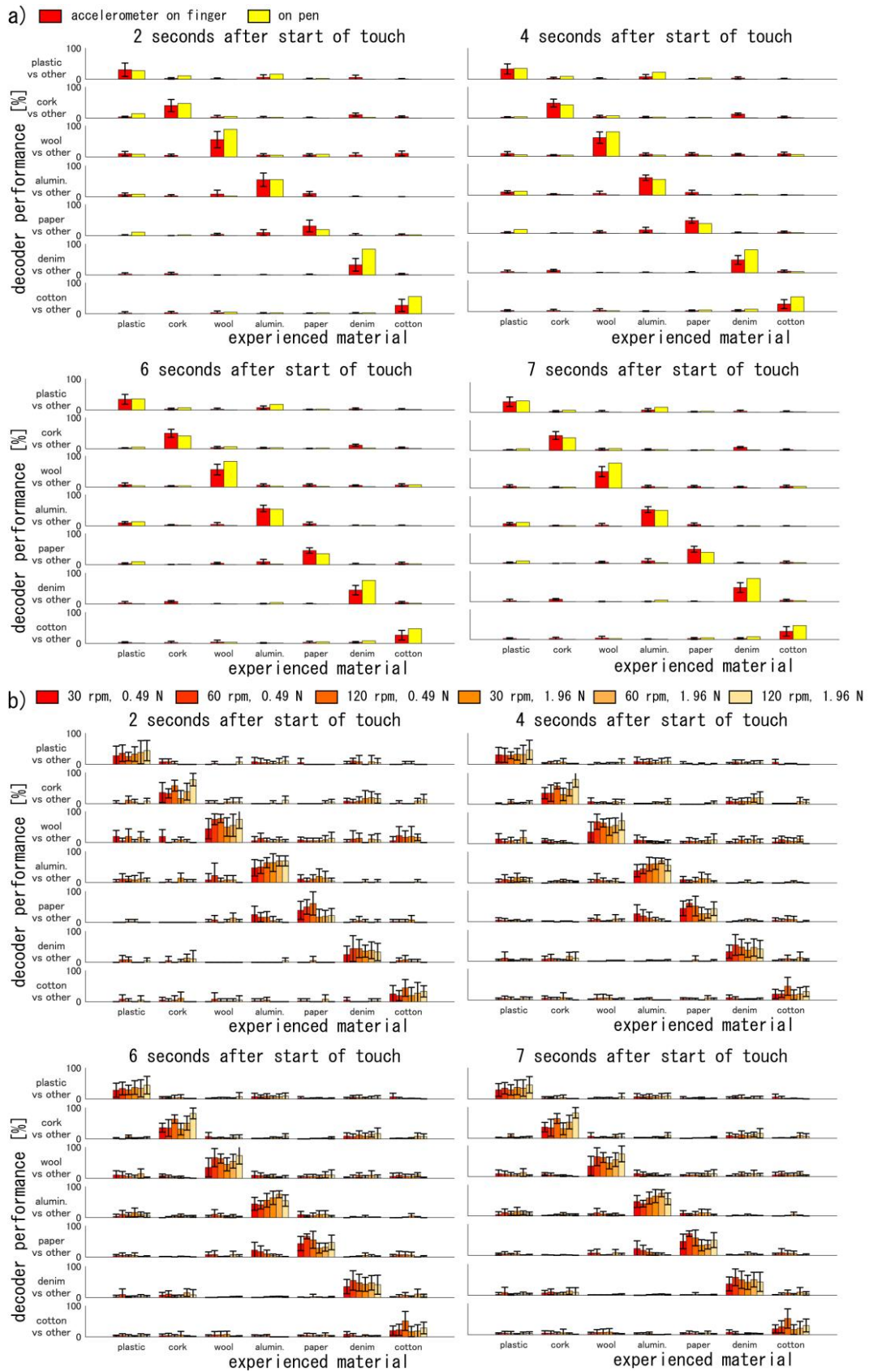

Figure 3. Performance of decoders: We utilized seven binary decoders, each to classify a data point as one of our test material (or not). a) shows the results of the classification by the seven decoders across all our participants (red data) and with the accelerometer mounted pen (yellow data). We observe that across participants and trials, all materials were classified correctly in the majority of the trials. b) shows the same data separated by the speed and load during the material touches.

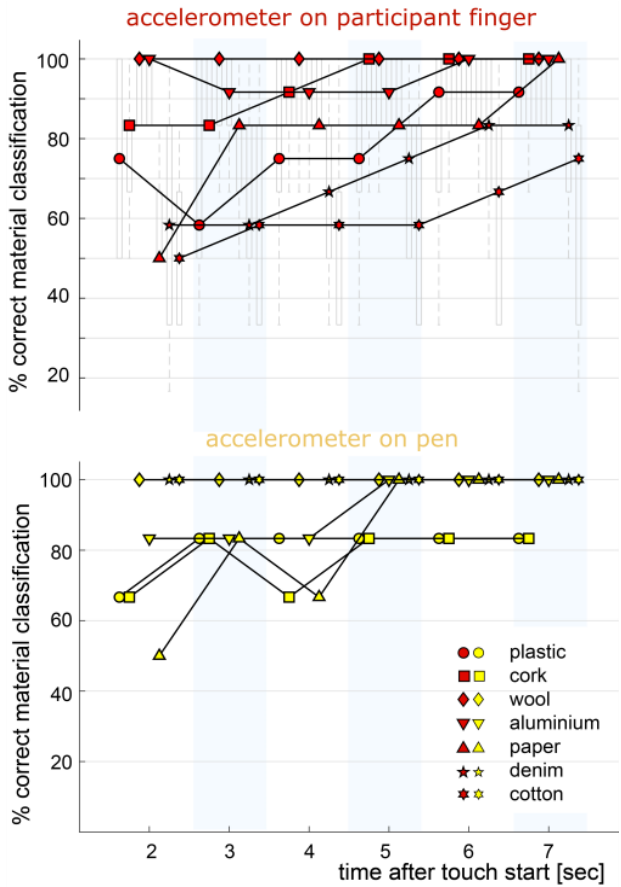

Figure 4. Trial wise performance. We adopted a winner take all approach to classify indiavidual touch trials. The figure shows the classification accuracy of each material across trials when the touches were performed by particpant fingers (upper panel) and by the pen (lower panel). The accuracy is plotted when using the first 2 (from first 1 to first 2) seconds, 3 (1 to 3) seconds, 4 (1 to 4) seconds, 5 (1 to 5) seconds, 6 (1 to 6) seconds, and 7 (1 to 7) seconds of data after the start of touch. The data points in the upper panel represent across particpant medians, the box edges represent the 25 and 75the percentile, while the whiskers represent the data ranges acros particpants.

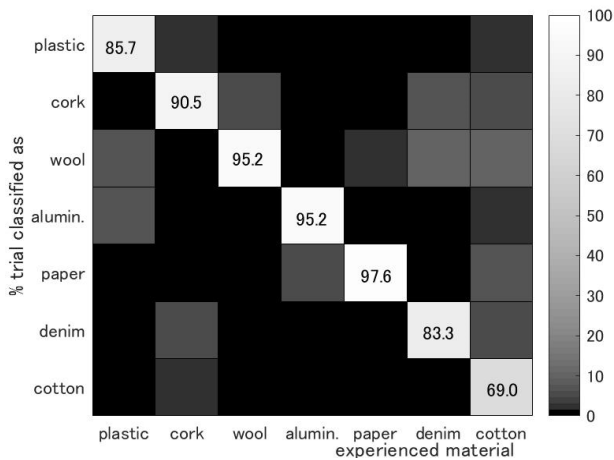

Figure 5. Confusion matrix of the clasification using the first 7 seconds of touch data. The figure has been plotted by combining the results for the particpant touches and the pen touches.

This procedure enabled us to understand how the amount of data (the is the length of touch) influenced the decoding results, while ensuring that the test and training data are distinct through our experiment.

*C. Integrating the seven decoder results for Texture decoding*

The data bins from each validation was classified by each of the seven decoders as being the material (corresponding to the decoder) or not. To make the final decision on the material being touched, we took a winner take all approach. That is, the touched sample was classified as the material into which most of the data bins in the trial were classified as.

### III. RESULTS

*A. Load force did not change with material*

We started by ensuring that the fingertip load force during our task was not modulated by the material touched, which could have in turn helped in the material decoding. We calculated the mean value of the fingertip force at each setting force (0.49, 1.96 N) for each material across subjects and conducted two-way ANOVA, across the speed and materials, to analyze the effect of texture on fingertip force. The ANOVA showed that there was no effect of material and speed on the load force either at the load setting of 0.49 N (effect of speed: $F(6, 60) = 1.71$, $p = 0.15$, effect of material: $F(2, 60) = 0.005$, $p = 0.99$, interaction: $F(12, 60) = 0.96$, $p = 0.49$) or 1.96 N (effect of speed: $F(6, 60) = 0.75$, $p = 0.69$, effect of material: $F(2, 60) = 0.63$, $p = 0.55$, interaction: $F(12, 60) = 0.96$, $p = 0.49$).

*B. Performance of the seven decoders across participants*

Fig. 2 shows the sample accelerometer recordings in three axes from one participant when he touched the different test materials in different trials. The average and standard deviation of the material decoding performance across the six participants is shown in Fig 3a in red for the decoding performed with different amounts of data. The yellow bar shows the same result when the material was touched with the accelerometer mounted pen. As explained earlier, the data is shown as a 2D histogram. Figure 3b shows the same data separated into the different load forces and speeds.

The diagonal heavy plots in Fig. 3 shows that across participants, if we consider 2 seconds or more of data, the materials were perfectly identified by the seven decoders in majority of the trials. The identification rates remain almost similar after the first 3 seconds. Hence, in order to preserve visibility, we show the data from only four test periods, using the first 2, 4, 6 and 7 seconds of data in Fig. 3. Furthermore,

Fig. 3b shows that the perfect identification occurred at every speed and load. Note that we do not input the touch speed and touch load to the decoders, which indicates that the decoders are able to recognize speed and load independent features of the materials in the accelerometer signals.

*C. Trial wise Material Classification*

Looking at Fig. 3, we decided to adopt a simple winner take all approach to decode the material in every trial. Fig. 4 plots the material classification result over individual trials across participants (upper panel), and with the pen (lower panel). The participant data is shown as box plots, with the median highlighted by markers corresponding to each material. Note that, as we classify between seven materials, the chance level in our experiment is 100/7=14.3 %.

First, we observe that within just 2 seconds of touch, the accelerometer signals can be used to sufficiently classify our seven materials above chance. When the touches are by the participant fingers, the classification accuracy quickly rises with time. By 6 seconds after touch, all the materials, except cotton, could be classified with over 80% accuracy, with five of the seven materials classified with over 90% accuracy. By seven seconds, except for denim (83.3%), and cotton (75.0%), all materials were classified with 100% median accuracy.

The performance is arguably better when the touch is made by the pen. Denim, cotton and wool were classified with an accuracy of 100% within 2 seconds of touch with the pen. Five of the seven materials could be classified with 100% accuracy within 5 seconds of touch with the pen. The remaining materials (plastic and cork) were classified with an accuracy of 83%. Overall cotton and denim were most difficult to classify when touched by a finger, while plastic and cork were most difficult to identify when touched with the pen.

Fig. 5 shows the confusion matrix when using data from seven seconds of touch (in Fig. 4). The figure has been plotted by combining the results from both the human fingers and pen. The figure shows that cotton and plastic were sometimes mis identified as wool or aluminum, while cork and denim were misidentified as each other. Overall, we see that the materials were classified with an accuracy of 88.1% across our experiment. If we omit cotton, this accuracy increases to 91.3% across the remaining six materials. Note that we are using just a low-cost accelerometer for the decoding- these results show that the signals from an accelerometer mounted on the finger, or a pen (our simulated prosthetic finger) can be used to decode the touched material with high accuracy.

## IV. DISCUSSION

In this study we tested whether an accelerometer mounted on a finger is sufficient for the identification of a touched material from the vibrations the touch induces. We observed that common materials, including plastic, cork, wool, aluminum, paper, denim, cotton can be decoded with about 88% accuracy using acceleration features alone.XXX remove reference 14.

Our method relies on the detection of vibrations induced by a touch on the finger. Therefore, a limitation of our method is that it is useful only when the touches are dynamic, that is when the finger is not static. However, considering the fact that accelerometers are increasingly cheaper to buy and smaller in size, we believe that, in spite of the requirement of movement, decoding of materials using accelerometers to be an attractive possibility for stroke rehabilitation and prosthetics technologies.

The good decoding performance with the accelerometer mounted on the pen (our simulated prosthetic finger) was in fact a pleasant surprise for us. We expected the decoding performance to be affected a lot be the absence of the compliance and finger prints (characteristic of a human finger) on the pen. However, we find that while some materials (specifically, plastic and cork) were better identified during touch by a human finger, other materials (specifically cotton and denim) were better identified when the touched by the pen. Further research is required to understand which finger material and shape is best for the identification of material using accelerometer recordings. The results of this research will be crucial for the design of prosthetics in the future.

A recent study suggested that combining force, vibration and surface temperature features can enable high decoding accuracy [17]. Taking a minimalistic approach, in this study, we have concentrated on acceleration features alone, again due to our motivation to develop in expensive and light decoder and due to the fact that while touch force and surface temperatures can be estimated when a prosthetic finger touches a surface, these are difficult to estimate in rehabilitation scenarios where the touches are made by a human hand (finger). Interestingly, we could achieve a good classification rate (that was better, and with more materials that [17]) just with the accelerometer signals. However, our texture analysis setup allows for the measurement of surface temperature, and we are now exploring whether and how the addition of surface temperature features to the decoder can further improve decoding performance.

Finally, we are also analyzing how the current accelerometer based on decoding can be improved with other decoding algorithms. Specifically, we are looking into two issues. First, (as mentioned above) in the current analysis we consider the accelerometer signals alone, without considering the touch speed and load force. We are now analyzing how decoding can improve if the touch speed and load force are available for the decoder to train on. Second, in the current study we used a very basic, winner take all approach to classify materials in trials. We are now considering other ways to integrate the results from our material decoders using popular Bayesian approaches like those used for haptic and visual classifications [18, 19].

In conclusion, in this our study we utilized an accelerometer fixed on the finger of participants and a simulated prosthetic finger, to show that acceleration features can provide very accurate identification of materials being touched by the finger across different speeds and touch loads. While there is still challenges to improve the results presented here, these results highlight the promising possibility of using accelerometers as low cost and light weight material/texture transducers.